\documentstyle[aps,prb,twocolumn,floats,epsf,rotate]{revtex}

\begin{document}
\twocolumn[\hsize\textwidth\columnwidth\hsize\csname
@twocolumnfalse\endcsname

\title{Skyrmion Dynamics and NMR Line Shapes in QHE Ferromagnets}

\draft

\author{Jairo Sinova$^{1}$, S.M. Girvin$^{1}$,T. Jungwirth$^{1,2}$,
and K. Moon$^{3}$}
\address{$^{1}$Department of Physics,
Indiana University, Bloomington, Indiana 47405}
\address{$^{2}$Institute of Physics ASCR,
Cukrovarnick\'a 10, 162 00 Praha 6, Czech Republic}
\address{$^{3}$Department of Physics, Yonsei University, Seoul 120-749, Korea}
\date{\today}
\maketitle

\begin{abstract}
The low energy charged excitations in quantum Hall 
ferromagnets are topological defects in the spin orientation 
known as Skyrmions. Recent experimental studies 
on nuclear magnetic 
resonance spectral line shapes in quantum well heterostructures
show a transition from a motionally 
narrowed to a broader ``frozen'' line shape as the temperature 
is lowered at fixed filling factor. 
We present a Skyrmion diffusion model that describes the experimental observations
qualitatively and shows a time scale of $\sim 50 \mu{\rm sec}$ for the
transport relaxation time of the Skyrmions. 
The transition is 
characterized by an intermediate time regime that we 
demonstrate is weakly sensitive to the dynamics of the charged spin
texture excitations and the sub-band electronic wave functions 
within our model. 
We also show that the spectral line shape is very sensitive to the nuclear
polarization profile along the $z$ axis obtained through  the optical 
pumping technique. 
\end{abstract}

\pacs{73.40.Hm, 76.60-k, 67.80.Jd, 73.20.Mf, 76.60.Cq}

\vskip2pc]
\section{INTRODUCTION}
In the presence of a strong magnetic field, a two-dimensional interacting electron gas 
(2DEG) exhibits many different quantum states depending on field strength, electron density,
and disorder. At Landau level filling factor $\nu=1$, when
the number of electrons is equal to the number of available Landau orbitals in
the lowest Landau level, the 2DEG is in an itinerant
ferromagnetic state where all the 
electron spins are aligned with the magnetic field.\cite{general}
The novel features
of this quantum Hall ferromagnet (QHF) state originate from the 
relative strength of 
the electron-electron Coulomb energy,
$e^2/\epsilon l_B$,
and the Zeeman energy, $g \mu_B B$, which are of order 
160 K and 3 K respectively in typical experimental situations. \cite{barret}
The Pauli exclusion principle combined with the overwhelming cost of Coulomb excitations
makes the single-particle
spin-$1/2$ electron excitation gapped at filling factor $\nu=1$.
The low energy (but still gapped) charge excitations of the system are 
Skyrmion spin textures containing many (4-30) flipped spins and are topologically
stable.\cite{general}
By paying the lower Zeeman price the spins can align locally creating a 
more advantageous charge distribution that lowers both the Hartree and
exchange Coulomb cost
with respect to a single spin flip excitation.
Because these excitations are the cheapest way to 
introduce charge into the system, they are present in the ground state of the system at
filling factors close to $\nu=1$.\cite{general,sondhi}
This produces a rapid reduction of the electron 
spin polarization away from $\nu=1$ which is observed in experiments. \cite{barret}

The probe of choice to study the electron spin polarization of the 2DEG is nuclear magnetic
resonance (NMR). Measurements of the Knight shift, which is linearly proportional to the
%
electron spin polarization, 
have unambiguously proven the existence of these exotic topological
charge excitations.\cite{barret} However, this rich probe 
keeps providing us with more surprising information about the Skyrmions, continuously 
challenging the theory of QHF.
Recent experiments carried out by N. N. Kuzma {\it et al}.,
\cite{yalegrouppapers} using optical pumping 
techniques to enhance the NMR signal of the 2DEG, have
focused on the NMR line shape as a function of temperature and filling factor near $\nu=1/3$
(which is also a QHF state). 
The observed free induction decay signals
show a large dependence of the electron polarization on filling factor, indicating
that charged spin excitations are more important near $\nu=1/3$ than previously
expected from theory.\cite{kamilla} 
Near this filling factor, as the temperature is lowered, 
the Knight shift increases with decreasing temperature 
(although having a local minimum)
until it reaches a saturation level at very low temperatures. The spectral
line shape ranges from the motionally narrowed regime, where the polarization shows a 
sharp peak due to the average polarization seen by each nucleus, to the 
frozen \cite{comment}
regime, where the peak is much broader due to the 
presence of frozen spin textures
in the ground state and the nuclei seeing different electron polarizations depending
on their location.  
The interesting regime is the intermediate one where the dynamical
time scale of the Skyrmions is comparable to the inverse frequency of the Knight shift.
This is the regime which contains the greatest wealth of information about Skyrmion 
dynamics. 
The situation is complicated by the fact that the Knight shift varies
strongly with position in the $z$ direction across the quantum well.
Nuclei at the edges of the well see a lower electron density and hence a
smaller Knight shift.  As a result, these nuclei can still be in the
motionally narrowed regime when nuclei in the center of the well are
already close to 
the frozen regime.  This subtlety is taken into account in
our analysis.  

Because the spin stiffness at $\nu=1/3$ is so small \cite{moon}
the Skyrmions will be very small and the continuum field theoretic
approach will be poorly controlled. We therefore focus on the 
case of filling factors near $\nu=1$ where measurements
are currently underway. \cite{barretnow}$^,$\cite{Sorin}
We present a simple model for the Skyrmion dynamics near $\nu=1$
in this paper.
We find that the spectral line shape, besides being 
dependent on the transport relaxation time 
of the Skyrmions, is very sensitive to the 
nuclear spin polarization density along the $z$ direction and,
to a lesser extent, the electron density profile in the
z-direction.

We organize this paper as follows. In Sec. II we introduce the theoretical
background needed for our model calculations. In Sec. III we present our
model and results. In Sec. IV we discuss the implications of the results
and possible new outlooks on this problem.

\section{THEORY}

In the free induction decay NMR experiments, after the nuclei in the wells 
are polarized by optical pumping,\cite{barret}$^,$\cite{opticalpumping}
the nuclear spins are tipped
by a $\pi/2$ pulse and allowed to precess freely. \cite{barret,yalegrouppapers}
These spins will precess at the
Larmor frequency produced by the local magnetic field. This local magnetic field is
composed of the external one plus a contribution from the electron polarization 
due to the Fermi contact hyperfine coupling which enters the nuclear spin 
Hamiltonian in the same way as the external magnetic field \cite{slichter}
\begin{equation}
{\cal H_{\rm N}}=-g_N \mu_N \sum_j{\bf S}_{j}\cdot 
\left( {\bf H}_{\rm o}+{\bf B}_e({\bf R}_j)\right)\,\,\, ,
\label{hamiltonian}
\end{equation}
where $\mu_N$ is the nuclear magneton, ${\bf S}_{j}$ is the nuclear spin 
in units of $\hbar$ at position ${\bf R}_j$,
${\bf H}_{\rm o}$ is the applied magnetic field,
and ${\bf B}_e({\bf R}_j)\equiv (-16 \pi \mu_B /3)\sum_i {\bf S}^e_i
\delta({\bf r}_i-{\bf R}_j)$  is the effective local magnetic
field contribution due to the electronic polarization with ${\bf S}^e_i$ 
and ${\bf r}_i$
being the spin and position of the $i$th electron. 

The last term in eq. (\ref{hamiltonian}) is the one responsible for the observed Knight 
shift.
For the purposes of computing the Knight shift, it is adequate to
replace ${\bf B}_e$ by its expectation value 
\begin{equation}
\langle B_z({\bf R},\nu,T)\rangle \propto |u({\bf R})|^2 {\cal
P}({\bf R},\nu,T)\,\,\, ,
\label{Beff}
\end{equation}
where $|u({\bf R})|^2$ is the electron envelope function obtained from a 
self-consistent local spin-density approximation calculation, and 
${\cal P}({\bf R},\nu,T)$ is the average electron spin polarization 
at position ${\bf R}$ for a given
filling factor and temperature. 
If we assume that the electron envelope function is only a function of $z$
(the growth direction)
we can further parameterize the local Knight shift as
$K_s(z)\equiv \rho_e(z) \tilde{K}_s {\cal P}({\bf R},\nu,T)$, where 
$\rho_e(z)=\int dx\,dy|u({\bf r})|^2$ is
the electron density along the z-direction (normalized to
unity at its maximum) 
and $\tilde{K}_s$ is a constant that can be fitted to the experimental
spectra at the lowest temperatures.

Next, we must connect these expressions to the observed intensity spectrum $I(\omega)$,
which is the time Fourier transform of the induced voltage produced in the 
tipping coil due to the precessing nuclear spins in the quantum wells. 
In the absence of in-plane spin nuclear decay, for a given nuclear spin,
the time evolution of the spin's expectation value 
(relative to the evolution with zero Knight shift)     
is given by 
\begin{eqnarray}
\langle S^+_j(t)\rangle &\equiv&
\langle S_{jx}+iS_{jy}\rangle \nonumber\\
&=&S_j^+(0) \exp[-i \int_0^t d\tau \tilde{K}_s \rho_e(z_j)
{\cal P}({\bf R}_j,\nu,T,\tau)]\,\,\, ,\nonumber
\end{eqnarray}
where now we have allowed the electronic polarization
to vary with time. \cite{slichter,note1}
The induced voltage is proportional to 
the time derivative of ${\rm Re}\left[\langle 
S^+_j(t)\rangle\right]$, which
is approximately $-{\rm Im}\left[ i\omega_0 \langle S^+_j(t)\rangle\right]$
whenever $\omega_0\gg \tilde{K}_s$, with $\omega_0$ being the bare nuclear
precession rate.
However, for nuclear spins in 
a solid the in-plane magnetization decays as a gaussian due to
nuclear 
dipole-dipole interactions.
This yields the final expression for the intensity
\begin{equation}
I(\omega)\propto \int d {\bf r} \rho_{\rm N}({\bf r}) {\rm Re}\left[\int_0^\infty
dt e^{-\sigma^2 t^2/2+i \omega t} \langle S^+({\bf r},t)\rangle\right]
\,\,\, ,
\end{equation}
where $\rho_{\rm N}({\bf r})=\langle\sum_j \delta ({\bf r}-{\bf R_j})
{S}_{z\,j} \rangle$ is the polarization density of 
nuclear spins (not the
number density of nuclei), and $\sigma=\Delta \omega/(2 \sqrt{2\ln{2}})$
with $\Delta \omega$ being the full width half maximum (FWHM) of the
unshifted NMR signal. In the samples used in the experiments
$\Delta \omega=2\pi 3.5 {\rm kHz}$, and therefore $\sigma=9.34 {\rm msec}^{-1}$. 
Any model describing the observed spectra has to contain realistic
estimates of ${\cal P}({\bf R}_j,\nu,T,t)$, $\rho_{\rm e}({\bf r})$,
and $\rho_{\rm N}({\bf r})$. Of these three the first two are the ones that 
have been studied most extensively. 
\cite{sondhi,moon,moon2,rajaraman}$^-$\cite{cote and allan}
In a previous model presented by Kuzma {\it et al}. \cite{yalegrouppapers}
to describe the transition near $\nu=1/3$, a two polarization
model for ${\cal P}(\nu,T,t)$ was used, and $\rho_e$ and $\rho_N$
were approximated by sinusoidal shapes. Here we attempt to improve 
upon this model by carefully examining the approximations used 
for each quantity.

At zero temperature
${\cal P}({\bf R}_j,\nu)$, the electron polarization, is exactly unity at
$\nu=1$ as mentioned in section I. However, as the filling factor goes away from
from $\nu=1$, Skyrmions begin to appear in the ground state of the
system. These charged spin texture excitations reduce the polarization
locally and a realistic approximation for the shape of the Skyrmion is
needed to obtain a reasonable ${\cal P}({\bf R}_j,\nu)$.
There have been many studies on the shape of the Skyrmions.
\cite{sondhi,moon,moon2,rajaraman}$^-$\cite{cote and allan}
Most analytical
approaches describing the Skyrmion excitations have taken the route
of effective field theories such as the modified O(3) non-linear sigma
model (NL$\sigma$),
\cite{rajaraman} where a Zeeman term and a Coulomb interaction term are inserted
in the classical NL$\sigma$ model. 
\cite{sondhi,moon,skyrmionO3model}
The magnetization profile of a Skyrmion obtained from this theory,
in the limit of zero Zeeman energy,
is given by
\[m_z({\bf r})=\frac{r_\perp^2-\lambda^2}{r_\perp^2+\lambda^2}\,\,\, ,\]
where $\lambda$ determines the size of the Skyrmion, and
$r_\perp$ is the projection of ${\bf r}$ on the plane (denoted by
$r$ henceforth). 
At larger distances away from the Skyrmion's center the Zeeman
term dominates and the magnetization goes as
$1-m_z({\bf r})\sim e^{-2\kappa r}/r$, with $\kappa=4.4 l_B$ near
$\nu=1$. \cite{skyrmionO3model}
This analytical approach has been very successful in explaining
qualitatively the physics of the QHF, however it has
not been able to predict quantitatively the shape of the Skyrmion 
excitations at experimentally accessible parameters.
In GaAs/Al$_x$Ga$_{1-x}$As heterostructures, at around 10 T, the
number of spin flips, $K$, per unit charge introduced near $\nu=1$ is \cite{barret}
$\sim 3-4$. Single Skyrmion microscopic calculations using
techniques such as Hartree-Fock\cite{allan} (HF), exact diagonalization,
and variational wave functions,\cite{abolfath} have been more successful
at obtaining quantitative agreement with experiments. These, however, are not as
transparent as the classical model in describing the physics behind the 
excitations. Also, Brey {\it et al}. \cite{cote and allan} 
have performed HF calculations 
of the skyrme crystals
formed at filling factors near $\nu=1$ which most
accurately describe the spin polarization observed in experiments.
\cite{barret,cote and allan}
The failure of the NL$\sigma$
model to predict the correct Skyrmion shape for small $K$ originates
in the truncation of the gradient expansion. The HF calculations 
are in essence self-consistent mean field calculations with the order of the
gradient expansion taken to infinity and hence are more successful at 
predicting the small Skyrmions that change shape on a much shorter length scale.
\cite{abolfath}
As described in the next section, we use a phenomenological form 
for $m_z(r)$ fitted to
the Hartree Fock calculations.

\begin{figure}
\epsfxsize=3.375in
\centerline{\epsffile{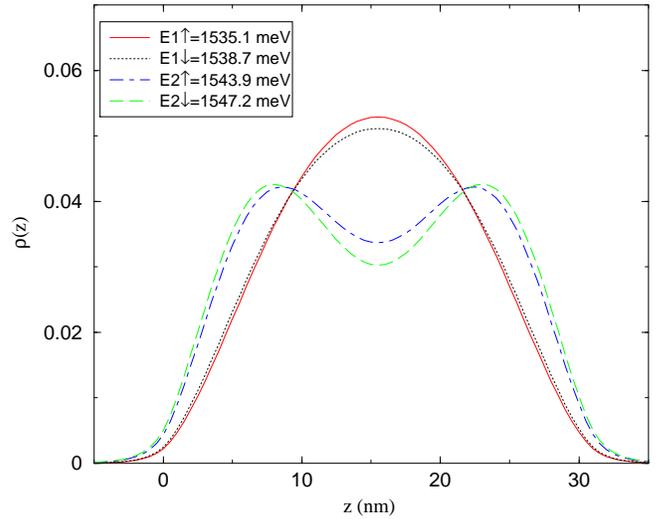}}
\caption{Electronic density using the local
spin-density approximation 
for a sample with $B_{total}=7.05{\rm T}$, a well width of $30 {\rm nm}$,
and a tilt angle of $\theta_{tilt}=28.5^o$. The energies are measured from the
top of the highest valence band.  
Note that the tilted magnetic field
smears out the locations of the node positions for states in the first
excited subband.
}
\label{elecdens}
\end{figure}

The electron density along the z-direction $\rho_{\rm e}(z)$, is obtained
from the electronic spin-split energy levels
of the GaAs heterostructures in the presence of a strong tilted field
(as is used in the experiments).
These charge distributions are calculated using a local spin-density
approximation. The density profiles of the two spin states
and two lowest levels,
measured with respect to the valence band for a $30-{\rm nm}$-wide GaAs single
quantum well,
are shown in Fig. \ref{elecdens}.
The parameters have been chosen to match the experiments \cite{barret} done near
$\nu=1$. 

The nuclear polarization profile estimate $\rho_N$, comes
from considering the experimental set up. After a train of rf pulses
which destroys the thermally induced nuclear polarization, the 
sample is radiated with circularly polarized $\sigma^+$ light
tuned to the band gap in the well. This excites 100\% polarized electrons
to the lowest unoccupied level (spin down)  and in the recombination process
the nuclear system absorbs part of the angular momentum transferred to the sample.
Hence, the optical pumping can produce an enhancement  in the nuclear polarization
by as much as a factor of 100.\cite{barret} This optical pumping is necessary,
at least at higher temperatures, to make the NMR signal visible. The
problem is that, even though the initial excitation process is understood, the
recombination process is much more complicated. In such a process, Skyrmions
and anti-Skyrmions are being created constantly and this may have an effect
on the nuclear polarization profile which is not well understood at present. 

One may ask why at such low temperatures
(1.6 K) and being so deep in the insulating phase of the quantum
Hall state one can even observe motional narrowing of the
NMR line shapes. We can answer this question in reverse. Given
the information from the experiments,\cite{barret,yalegrouppapers}
what longitudinal resistivity can we infer? The 
onset of the frozen regime indicates that $D K_s^{-1}\sim n^{-1}_{skyr}$
and $dn/d\mu\sim n/\Delta E$, with D being the diffusion 
constant, $n_{skyr}$ and $n$ the density of Skyrmions and electrons
respectively, and $\Delta E$ the disorder broadening of the
Landau level which can be estimated to be at least of the order of
$\sim 10 $K and is possibly much larger. \cite{DasSarma}   
This information can be inserted in the Einstein relation for
the conductivity
\begin{eqnarray}
\sigma_{xx}&=&e^2 D \frac{dn}{d\mu}\sim e^2 n_{skyr}^{-1}
\frac{K_s n}{\Delta E}\\&\sim&
\frac{3\times 10^{-12} \Omega ^{-1}}{|\nu-1|}\,\,,
\end{eqnarray}
which gives 
a lower bound on the inverse conductivity 
$1/\sigma_{xx}\sim 300 \,{\rm G}\Omega |\nu-1|$. Hence, we see that 
the dynamics of the Skyrmions can appear to be fast on the
NMR time scale even deep in the insulating regime.

\section{THE MODEL AND RESULTS}
When calculating the spectral intensity from eq. (3) it is useful to 
first think about the different time 
scales in the problem and how relevant each one is in calculating the
NMR spectral line shape. At the experimental fields used
(7.05 T) the bare nuclear precession rate is of order 100 MHz, the 
extra precession rate created by the electron spins (the Knight shift) is of 
order 20 KHz, the nuclear spin-lattice relaxation rate ranges from
4 mHz to 45 mHz, and the in-plane spin relaxation rate due to the nuclear dipolar
coupling is approximately 3.5 KHz.\cite{barret}
Hence, in the calculations that follow,
we shall ignore the spin-lattice relaxation rate and place our zero
of frequency at the bare NMR resonance. 
Furthermore, since we assume perfect cubic symmetry of the GaAs crystal, we
omit any effects on the spectral calculation
due to any nuclear quadruple splitting. 

Rather than doing a full microscopic
calculation for
${\cal P}({\bf R},\nu,T)$
in the presence of disorder, we take a more modest aim and focus
on the transition where the spectral line shape goes from the frozen
regime to the motionally narrowed regime. For $|\nu-1|\ll 1$ the ground
state of the 2DEG at $T=0$ is believed to be in a skyrme square lattice
state. \cite{cote and allan} In practice, the skyrme lattice is
melted in most of the accessible temperature ranges. \cite{skyrmionO3model,cote}
However, at the low experimental temperatures considered here
(T$< 4$ K), although the long length scale correlations
vanish, we expect the short length scale correlations
to contain crystal-like features.
To model this
we introduce a Skyrmion square lattice
with the unit cell size given by the appropriate filling factor.
We take the magnetization profile of the Skyrmion to be:
\[m_z(r)=\frac{r^2-\lambda^2 e^{-\alpha\frac{r^2}{L^2}}}{
r^2+\lambda^2 e^{-\alpha\frac{r^2}{L^2}}} \,\, ,\]
which, by choosing $\lambda$ and $\alpha$ appropriately, can resemble
closely the HF calculations previously done.\cite{allan}
For $\nu=0.96$ we used $\alpha=5.6$ and $\lambda^2=2.6$.
Although in the HF calculations $m_z(0)\approx-0.6$ 
due to the zero point
fluctuations, we find that this has little effect on the spectral
line shape,
and we hence keep our simpler
functional form in the model calculations. 
Also, in any experimental situation, the
part of the spectrum  due to this small region of fully reversed spins
tends to be weak.
To model the time dependence of ${\cal P}({\bf R},\nu,T,t)$, instead
of allowing the Skyrmions to undergo correlated
thermally induced motion about their lattice 
points, we make the whole lattice move together to simplify the
numerical calculations. 
Hence, the
lattice is only allowed to move collectively in a random walk 
by performing a jump of average distance $l$ with a probability
$dt/\tau_J$ in the time interval $t$ to $t+dt$. The diffusion time 
across a unit cell of size
$L$,  $\tau_{\rm diff}=({L^2}/{l^2}) \tau_J$, 
is held constant for a given temperature and $\tau_J$ and $l$ are
 varied 
to test the sensitivity of the spectral line shape to the microscopic
details of the dynamics. The motionally 
narrowed regime occurs when $\tau_{\rm diff}\ll K_s^{-1}$ and each 
nucleus experiences the average electron polarization in the sample. In the
limit $\tau_{\rm diff}\gg K_s^{-1}$ the Skyrmions are spatially frozen 
on the experimental time scale, 
with random motion being rare. This frozen regime of
the spectral line shape, observed at the lowest temperatures,
contains information on the actual shape and static distribution
of the Skyrmion excitations and the nuclear density profile.
The intermediate regime, $\tau_{\rm diff}\sim K_s^{-1}$, occurs when 
the FWHM of the spectrum reaches its peak value.
In this regime 
we find the greatest sensitivity of the line shape to the choice of
$\tau_{\rm J}$ for a fixed $\tau_{\rm diff}$. 

\begin{figure}
\epsfxsize=3.375in
\centerline{\epsffile{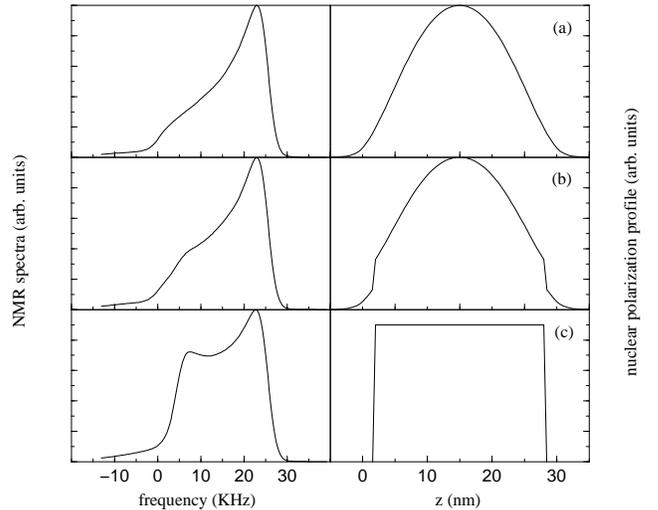}}
\caption{NMR spectra (on the left) at $\nu=1$ 
for different nuclear polarization
profiles (on the right). Profile (a) is taken from
the lowest unoccupied spin-split energy levels (spin down), profile (c)
is a constant polarization density truncated
at the edges, and (b) is a combination of (a)
and (c).} 
\label{nuc_sens}
\end{figure}

The adjustable parameter $\tilde{K}_s$, is
fixed by fitting the peak frequency of the spectral line shape to the one
observed experimentally in the frozen regime. The temperature is gauged
by $\tau_{\rm diff}$ and {\it calibrated} by the onset of the 
motionally narrowed and frozen regimes observed in the experiments.
We approximate the nuclear magnetization density
by the electronic density 
from the lowest unoccupied 
level (E1$\downarrow$
in Fig. 1), since the induced polarization
due to optical pumping is proportional to the local electronic density
of the electrons excited in such process. \cite{opticalpumping}
This gives a much more accurate profile
than the ones approximated by simple sinusoidal shapes. 
\cite{yalegrouppapers}
One striking result of our model is the high sensitivity of the
spectral line shape to the nuclear magnetization density. This 
sensitivity is illustrated in Fig. \ref{nuc_sens} where we show
different spectral line shapes at $\nu=1$ for several
nuclear polarization profiles.
This result strongly suggest that the details of the line
shape will be difficult to understand from first principles
without a better microscopic understanding of the optical
pumping process and its effect on the nuclear polarization
profile, together with other processes that may 
be affecting this profile. 

\begin{figure}
\epsfxsize=3.375in
\centerline{\epsffile{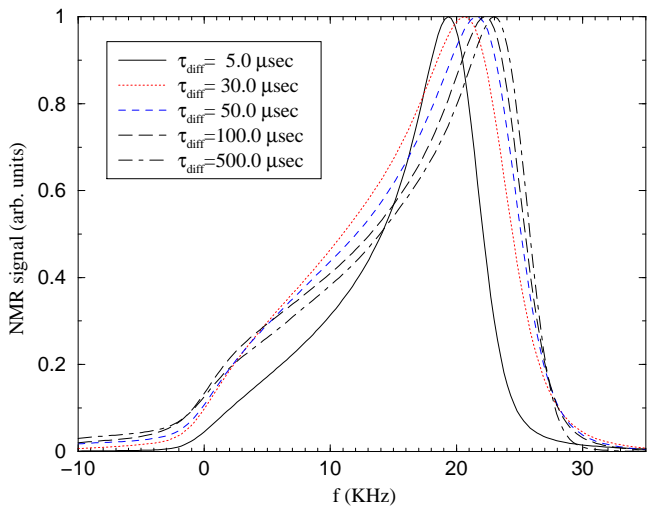}}
\caption{NMR spectra at $\nu=0.96$ for different diffusion times.
Here the parameters used are $\lambda^2=2.6$ and $\alpha=5.6$.
Note that here the global electron polarization, $\bar{\cal P}$, is fixed, and therefore
the peak location is no longer an accurate measure of $\bar{\cal P}$.} 
\label{spectra2}
\end{figure}

In Fig. \ref{spectra2} we show the spectral line shapes at different
diffusion times (temperatures). The maximum FWHM is obtained between 
40 and 50 $\mu$s. Also, note that $K_s$, which corresponds
to the maximum in the spectra, decreases monotonically as 
the temperature (or $\tau_{\rm diff}^{-1}$) increases. 
It is important to note that 
at temperatures where the Skyrmion
dynamics begin to `freeze' in 
the NMR time scales, the peak of the
spectrum (what is 
\noindent usually understood by the Knight shift) is no longer a good
measure of the global average electron polarization,
$\bar{\cal P}\equiv \int d^2r {\cal P}(r,\nu,T,t=0)$. Instead, in order to avoid
dynamic effects, one should measure the first moment of the spectra when
measuring $\bar{\cal P}$, since
\begin{figure}
\epsfxsize=3.375in
\centerline{\epsffile{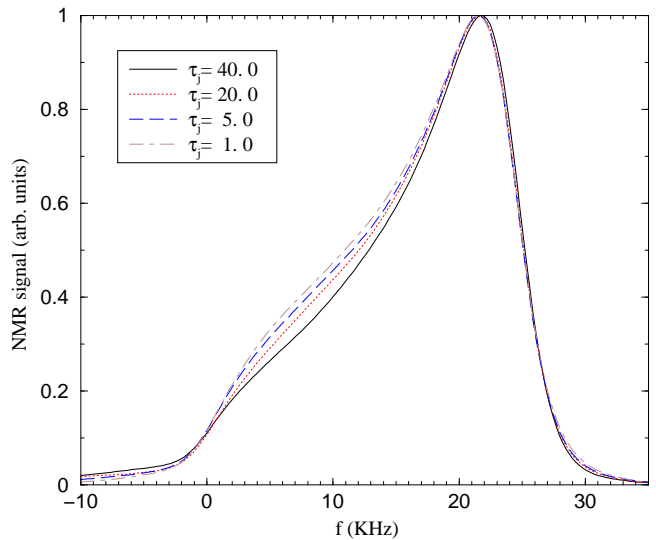}}
\caption{NMR spectra for $\tau_{\rm diff}=40 \mu{\rm sec}$ and
different $\tau_J$'s. Here $\nu=0.96$, $\lambda^2=2.6$, and $\alpha=5.6$.} 
\label{spectra1}
\end{figure}

\begin{equation}
\int_{-\infty}^{\infty}d \omega\,\omega I(\omega)\propto
\left[\int dz \rho_N(z) \rho_e(z)\right] \tilde{K}_s \bar{\cal P}\,,
\end{equation} 
where here we have $\int_{-\infty}^{\infty}d \omega\,I(\omega)=1$
rather than $I_{\rm max}=1$ as used in Fig. 3 and 4.

The sensitivity of 
the line shape to the choice of $\tau_J$
at $\tau_{\rm diff}\sim
40-50 \mu {\rm s}$ (corresponding to the maximum FWHM) is 
shown in Fig. \ref{spectra1}. This sensitivity, although weak, is
completely absent
at other $\tau_{\rm diff}$'s in the other regimes.

\section{CONCLUSION}

Our model illustrates qualitatively the behavior of the 
NMR spectra as a function of temperature for $T<4$ K.
It accurately predicts
a peak in the FWHM as a function of temperature.
It also shows a monotonic increase of the Knight shift
with decreasing temperature, reaching a plateau at
the lowest temperatures (largest $\tau_{\rm diff}$),
however, it does not
reproduce the local minimum in the Knight shift as a function of temperature
observed near $\nu=1/3$. \cite{yalegrouppapers}
Since the spin stiffness near $\nu=1/3$ is very small, we
do not expect this model to be valid near such filling factor,
where the Skyrmions are small and not well understood as is
the case near $\nu=1$.

We have also demonstrated that a full understanding of the NMR
line shape must involve a better understanding of the 
nonequilibrium nuclear polarization profile. This profile
is affected primarily by the optical pumping and possibly
by other thermal relaxation processes. 
We also have shown that to measure the behaviour of the 
average electron polarization $\bar{\cal P}$ at these 
temperatures, one must measure the first moment of the intensity
spectrum rather than its peak. 
We emphasize that our model does not attempt to calculate the
highly quantum-mechanical motion of the Skyrmions. This motion
may involve a semi-classical percolation in the case of
heavy nonlocalized Skyrmions
or variable range hopping in the case of highly
localized Skyrmions in a random potential. Our phenomenological
model does capture the appropriate time scales in the experiments
which should be an essential result of a more sophisticated
treatment of the problem.

The authors would like to thank Ren\'e C\^ot\'e 
and M. Abolfath for providing
the HF data for comparison, and Allan MacDonald,
S.E. Barrett, S. Melinte, V. Bayot, and S. Das Sarma for helpful discussions.
This work was supported by
grant NSF DMR 9714055, by grant 
INT-9602140, by the Ministry of Education of the Czeck Republic
under grant ME-104 and by the Grant Agency of the
Czeck Republic under grant 202/98/008J.


\begin{references}
\bibitem{comment} Here ``frozen" is understood as having a dynamical scale much
longer than the one in the experiments.
\bibitem{note1} Here we assume the spin lines up along the x-direction after the 
$\pi/2$ pulse. Also, we are neglecting any transient in-plane electronic polarization
which can be shown to give a negligible contribution to the spectral line shape.
\end{references}
\end{document}